\documentclass[copyright]{eptcs}
\providecommand{\event}{T. Neary, D. Woods, A.K. Seda and N. Murphy (Eds.):\\ The Complexity of Simple Programs 2008.}
\providecommand{\volume}{1}
\providecommand{\anno}{2009}
\providecommand{\firstpage}{199}
\usepackage{breakurl}

\usepackage{amsthm,graphicx,amssymb}

\newtheorem{thm}{Theorem}
\newtheorem{cor}[thm]{Corollary}
\newtheorem{opb}[thm]{Open Problem}
\newtheorem{defn}{Definition}

\title{Intrinsically Universal Cellular Automata}
\author{Nicolas Ollinger
    \email{Nicolas.Ollinger@lif.univ-mrs.fr}
    \institute{Laboratoire d'informatique fondamentale de Marseille (LIF),\\
        Aix-Marseille Universit\'e, CNRS,\\
        39 rue Joliot-Curie, 13\kern 0.2em 013 Marseille, France}
}
\def\authorrunning{N. Ollinger}
\def\titlerunning{Intrinsically Universal Cellular Automata}

\begin{document}
\maketitle

\begin{abstract}
This talk advocates intrinsic universality as a notion to identify simple cellular automata with complex computational behavior. After an historical introduction and proper definitions of intrinsic universality, which is discussed with respect to Turing and circuit universality, we discuss construction methods for small intrinsically universal cellular automata before discussing techniques for proving non universality.
\end{abstract}


\section{Introduction}

Universal machines are, in some way, the simplest type of complex machines with respect to computational aspects: the sum of all possible behaviors. Universality is also a convenient tool in computation as a way to transform data, that can be further manipulated by the machine, into code. Therefore, computation universality is one of the basic ingredients of self-reproducing cellular automata first introduced by von~Neumann~\cite{Neumann:1966:TSR} and in the subsequent works of Codd~\cite{Codd:1968:CA} and others\footnote{although some later constructions are not universal} to achieve construction universality. Since then, universality has been studied for itself both in the case of two-dimensional and one-dimensional cellular automata. For a detailed historical study, see the survey~\cite{NO:2008:JAC}.

\medskip
In the 60s and 70s, universality was mainly studied for high-dimensional cellular automata (2D, 3D). In this context, it seems natural, to achieve universality, to take inspiration from real-world computers by simulating components of boolean circuits. Wires, gates, clocks, fan-out, signal crossing, \emph{etc} are embedded into the configuration space of some local rule. Using these components, under the assumption that the family of elements is powerful enough, one obtains a universal cellular automaton under every reasonable hypothesis: from boolean circuits, one can wire finite state machines and memories to simulate sequential machines like Turing machines; or, one can wire finite state machines encoding the local rule of a cellular automaton and put infinitely many copies of that machine on a regular lattice, using wires to connect and synchronize the grid of automata to simulate the behavior of the encoded cellular automaton. This way, Banks~\cite{Banks:CA,Banks:1971:IPT} was able to construct very small universal 2D cellular automata.

\medskip
In the 70s and 80s, the study of cellular automata shifted to the one-dimensional space, motivated by the formal study of parallel algorithmics and formal languages recognition. In 1D, boolean circuits are no more a natural tool, but, as a configuration looks like a biinfinite tape, simulation of sequential machines like Turing machines is straightforward and provides the basis for a notion of computational (that is Turing) universality. This approach was developed by Smith~III~\cite{Smith:1971:SCU}. A major difficulty with Turing universality is the lack of a formal precise and general definition. The problem arises from two sources. First, a good commonly accepted formal definition of universality for Turing machines does not seem to exist. Second, encoding finitely described Turing machines configurations into infinite configurations, giving a reasonable halting condition, and a decoding of the result, is a delicate task. For a discussion on this formalization problem, see the study by Durand and R\'{o}ka~\cite{Durand:1999:GLU} of the universality of Conway's \emph{Game of Life\/} \cite{Berlekamp:1982:WWF}.

\medskip
In 1D, one can also consider simulating the cells of a configuration of a simulated cellular automaton by blocks of cells of a configuration of a simulator cellular automaton, leading to a notion of intrinsic universality. This notion, that coincide with the notion of boolean circuit universality in the case of 2D cellular automata, was first pointed out in the one-dimensional case by Banks~\cite{Banks:CA,Banks:1971:IPT} in the conclusion of its 2D construction, then rediscovered by Albert and \v{C}ulik~II~\cite{Albert:1987:SUC}. An attempt of a formal definition was given in Durand and R\'{o}ka~\cite{Durand:1999:GLU}. Whereas intrinsic universality implies Turing universality, one can prove that the converse is false, see Ollinger~\cite{Ollinger:2002:PHD}. Intrinsic universality is the topic of this talk.

\medskip
The paper continues as follows. In section~\ref{sec:def}, proper definitions of cellular automata and two definitions of intrinsic universality are proposed together with the main structural results. In section~\ref{sec:cons}, the construction of small universal cellular automata is discussed. In section~\ref{sec:non}, the more difficult question of non universality is considered.

\section{Definitions and Properties}\label{sec:def}

A $d$-dimensional \emph{cellular automaton\/} is a tuple $(d,S,N,f)$ where $S$ is the finite set of states, $N\subseteq_{\mbox{\tiny finite}}\mathbb{Z}^d$ is the finite \emph{neighborhood}, and ${f : S^N\rightarrow S}$ is the \emph{local rule\/} of the cellular automaton. A \emph{configuration\/} $c\in S^{\mathbb{Z}^d}$ is a coloring of the space $\mathbb{Z}^d$ by $S$. The \emph{global function\/} ${G : S^{\mathbb{Z}^d}\rightarrow S^{\mathbb{Z}^d}}$ maps a configuration $c$ to its image $G(c)$ by applying the local rule synchronously and uniformly according to $N$, \textit{i.e.}, for all $z\in\mathbb{Z}^d$, $G(c)(z)=f(c_{\mid z+N})$. The set of configurations $S^{\mathbb{Z}^d}$ is endowed with the Cantor topology, \textit{i.e.}, the product topology over $\mathbb{Z}^d$ of the discrete topology on $S$. This topology is metric, compact, and perfect. Under this topology, continuity corresponds to locality, as clopen sets correspond to sets of all configurations having a finite pattern in a given finite set, \textit{i.e.}, if ${C\subseteq S^{\mathbb{Z}^d}}$ is a clopen, there exists $M\subseteq_{\mbox{\tiny finite}}\mathbb{Z}^d$ such that for all $c\in C$, $\left\{c'\in S^{\mathbb{Z}^d}\middle|c_{\mid M}=c'_{\mid M}\right\}\subseteq C$. Adding invariance by translation, one can enforce uniformity and characterize cellular automata. The \emph{translation}, or \emph{shift}, over $S$ with translation vector $p\in\mathbb{Z}^d$, is the map ${\sigma_p:S^{\mathbb{Z}^d}\rightarrow S^{\mathbb{Z}^d}}$, satisfying, for all $c\in S^{\mathbb{Z}^d}$ and $z\in\mathbb{Z}^d$, $\sigma_p(c)(z+p)=c(z)$.

\begin{thm}[Hedlund~\cite{Hedlund-1969-EAS}, Richardson~\cite{Richardson-1972-TLT}] A map is the global function of a cellular automaton if and only if it is a continuous map commuting with translations.	
\end{thm}

This theorem allows us to manipulate cellular automata by their global function, composing them, inverting bijective ones, taking cartesian products, \textit{etc} being sure that the result is still the global function of a cellular automaton. For a proper study of cellular automata and their properties, one can read the survey of Kari~\cite{Kari:Survey}.

\medskip
A cellular automaton $\mathcal{A}$ is a \emph{subautomaton\/} of a cellular automaton $\mathcal{B}$, denoted as $\mathcal{A}\sqsubseteq\mathcal{B}$, if there exists an injective map ${\varphi : S_\mathcal{A}\rightarrow S_\mathcal{B}}$ such that ${\overline{\varphi}\circ G_\mathcal{A} = G_\mathcal{B}\circ\overline{\varphi}}$, where $\overline{\varphi}(c)=\varphi\circ c$ for all configuration $c\in S_\mathcal{A}^{\mathbb{Z}^d}$. For all $m\in\left(\mathbb{Z}^+\right)^d$, $n\in\mathbb{Z}^+$ and $k\in\mathbb{Z}^d$, the \emph{$\left<m,n,k\right>$-rescaling\/} of a cellular automaton $(d,S,N,f)$ is the cellular automaton with global function
\[
G^{\left<m,n,k\right>} = b^{m}\circ \sigma_k\circ G^n\circ b^{-m}
\]
where ${b^m:S^{\mathbb{Z}^d}\rightarrow \left(S^{\prod m}\right)^{\mathbb{Z}^d}}$ is the bijective $m$-\emph{packing map} satisfying, for all $c\in S^{\mathbb{Z}^d}$, $z\in\mathbb{Z}^d$ and $\alpha\in\prod m$, the equation $b^m(c)(z)(\alpha) = c(mz+\alpha)$, and $b^{-m}$ is the inverse of $b^m$.

\medskip
The \emph{injective bulking} quasi-order $\leqslant_i$ on cellular automata is defined thanks to subautomaton and rescaling. A cellular automaton $\mathcal{A}$ is simulated by a cellular automaton $\mathcal{B}$, denoted $\mathcal{A}\leqslant_i\mathcal{B}$, if there exists two rescalings $\left<m,n,k\right>$ and $\left<m',n',k'\right>$ such that $\mathcal{A}^{\left<m,n,k\right>}\sqsubseteq\mathcal{B}^{\left<m',n',k'\right>}$. This relation is a quasi-order with interesting structural properties, see \cite{Ollinger:2002:PHD}. It admits a maximal equivalence class that captures the notion of intrinsic universality used in most constructions of the literature. Moreover, the maximal class $\mathcal{U}_i$ admits a stronger characterization.

\begin{defn} A cellular automaton $\mathcal{A}$ is \emph{intrinsically universal\/} with respect to injective bulking if, for all cellular automata $\mathcal{B}$, there exists a rescaling $\left<m,n,k\right>$ such that ${\mathcal{B}\sqsubseteq\mathcal{A}^{\left<m,n,k\right>}}$.
\end{defn}

\medskip
A cellular automaton $\mathcal{A}$ is a \emph{mixautomaton\/} of a cellular automaton $\mathcal{B}$, denoted as $\mathcal{A}\unlhd\mathcal{B}$, if there exists a map ${\phi : S_\mathcal{A}\rightarrow 2^{S_\mathcal{B}}}$ with disjoint images, \textit{i.e.}, such that for all $s,s'\in S_\mathcal{A}$, $\phi(s)\cap\phi(s')=\emptyset$, such that ${\overline{\phi}\circ G_\mathcal{A} \supseteq G_\mathcal{B}\circ\overline{\phi}}$.

\medskip
The \emph{mixed bulking} quasi-order $\leqslant_m$ on cellular automata is defined thanks to mixautomaton and rescaling. A cellular automaton $\mathcal{A}$ is simulated by a cellular automaton $\mathcal{B}$, denoted $\mathcal{A}\leqslant_m\mathcal{B}$, if there exists two rescalings $\left<m,n,k\right>$ and $\left<m',n',k'\right>$ such that $\mathcal{A}^{\left<m,n,k\right>}\unlhd\mathcal{B}^{\left<m',n',k'\right>}$. This relation is a quasi-order with interesting structural properties, see Theyssier~\cite{Theyssier:2005:PHD}. As injective bulking is a refinement of mixed bulking, mixed bulking admits a maximal equivalence class that captures the notion of intrinsic universality used in most constructions of the literature. Moreover, the maximal class $\mathcal{U}_m$ admits a stronger characterization.

\begin{defn} A cellular automaton $\mathcal{A}$ is \emph{intrinsically universal\/} with respect to mixed bulking if, for all cellular automaton $\mathcal{B}$, there exists a rescaling $\left<m,n,k\right>$ such that ${\mathcal{B}\unlhd\mathcal{A}^{\left<m,n,k\right>}}$.
\end{defn}

Until now, all cellular automata known to be universal for mixed bulking can be shown universal for injective bulking (even if it is sometimes technical and painful).

\begin{opb} Does $\mathcal{U}_i=\mathcal{U}_m$?
\end{opb}

For a proper study of bulkings, there motivation and structural properties, see Delorme~\textit{et al}~\cite{UBULKI,UBULKII}. In the following, we will always consider injective bulking and the class $\mathcal{U}_i$... but everything remains true if you replace injective by mixed.

\medskip
A nice property of intrinsic universality is that its formal definition captures the constructions of the literature and provide a tool to prove non universality. A first natural question is to discuss decidability. Given a cellular automaton, can we decide if it is universal?

\begin{thm}[Ollinger \cite{NO:2003:IUP}] Intrinsic universality is undecidable.
\end{thm}

This theorem is proved by reducing the nilpotency problem of cellular automata on periodic configurations, obtained by Mazoyer and Rapaport~\cite{Mazoyer:1999:GFP}. As expected, there is no automatic method to test whether a cellular automaton is intrinsically universal.

\section{Constructing Small Universal Cellular Automata}\label{sec:cons}

Intrinsic universality is a recursively enumerable property. To prove that a given cellular automaton is universal, it is sufficient to prove that it simulates a fixed intrinsically universal cellular automaton. The undecidability comes from the fact that the size of the blocks needed to simulate one cell can grow unrecursively large with respect to the given cellular automaton.

\medskip
An intrinsically universal cellular automaton has to simulate cells: an entity computing a local rule and transmitting information to its neighbors. It is a straightforward exercise to construct an intrinsically universal cellular automaton with a small neighborhood and less than twenty states. For a general technique and examples, see \cite{NO:2008:JAC}. How to optimize the number of states with respect to a fixed neighborhood?

\medskip
In dimension 2, with von Neumann neighborhood $\{(0,0),$ $(1,0),$ $(0,1),$ $(-1,0),$ $(0,-1)\}$, encoding boolean circuits, Banks was able to construct a 2 states universal automaton. Notice that finite configurations are encoded into ultimately periodic configurations by intrinsically universal cellular automata.

\begin{thm}[Banks~\cite{Banks:CA,Banks:1971:IPT}] There exists an intrinsically universal 2D cellular automaton with von~Neumann neighborhood and 2 states.	
\end{thm}

\medskip
In dimension 1, a first technique, by Banks, consists of transforming a given universal cellular automaton of dimension 2 by having it simulating one dimensional cellular automata on a torus. The price to pay is either an extended neighborhood (neighbors are far from each other) or an extended set of states.

\begin{cor}[Banks~\cite{Banks:CA,Banks:1971:IPT}] There exists an intrinsically universal 1D cellular automaton with 2 states and a neighborhood of size 5.
\end{cor}

\medskip
Even if boolean circuits simulation is difficult, it can still be achieve in dimension 1. By a careful design, we were able in \cite{Ollinger:2002:TQF} to construct a universal automaton with 6 states and first neighbors neighborhood ${\left\{-1,0,1\right\}}$. Using particles and collisions, Richard was able to design a better simulation of one-way cellular automata and obtain the smallest know intrinsically universal cellular automaton, with only 4 states!

\begin{thm}[Ollinger and Richard~\cite{Richard:4st}] There exists an intrinsically universal 1D cellular automaton with first neighbors neighborhood and 4 states.
\end{thm}

\medskip
In the realm of Turing universality, the smallest universal automaton in dimension 1 has 2 states:

\begin{thm}[Cook~\cite{Cook:110}] Rule 110 is Turing universal.
\end{thm}

The construction makes heavy use of particles and collisions to simulate special variation of tag systems, see Richard~\cite{Richard:110} for a formal proof using bidimensional tools for particles and collisions (as used for the 4 states automaton). Unfortunately, a careful study of the set of collisions used show that, contrary to what is claimed by Wolfram~\cite{Wolfram:NKOS}, the construction cannot be used to prove intrinsic universality of rule 110.

\begin{opb} Is rule 110 intrinsically universal?
\end{opb}

\section{Identifying Non Universal Cellular Automata}\label{sec:non}

Thanks to the formal definition, a proof of non universality is possible for given cellular automata. However, due to the undecidability of the property, there is no general method. Still, several invariants can be used, coming from the study of the bulking quasi-orders.

\medskip
A first interesting problem is the pattern problem. Given a cellular automaton, a finite configuration and a pattern, the question is to decide if the pattern appears in the orbit of the configuration. For Turing universal, thus also for intrinsically universal cellular automata, the problem has maximal complexity. For simpler cellular automata it is decidable.

\begin{thm} The pattern problem is undecidable for any intrinsically universal cellular automaton.
\end{thm}

\medskip
A second interesting problem is the verification problem. Given a cellular automaton, a finite pattern and a state, the question is to decide if the state is obtained by iterating the local rule on the pattern until it consists of only one state. A simple simulation permits to test this property in polynomial time. By adapting the classical proof of Ladner~\cite{Ladner:1975}, one can prove that it is P-complete in general. For simpler cellular automata, the complexity can be lower.

\begin{thm} The verification problem is P-complete for any intrinsically universal cellular automaton.
\end{thm}

\medskip
Unfortunately, in the case of rule 110, the pattern problem is undecidable and the verification problem has been shown P-complete by Neary and Woods:

\begin{thm}[Neary and Woods~\cite{Woods:P-complete}] Rule 110 is P-complete.
\end{thm}

\begin{opb} Develop more tools to prove non universality.
\end{opb}

\bibliographystyle{eptcs}

\end{document}